\documentclass[sigconf]{acmart}

\usepackage{balance}
\usepackage{subcaption}

\AtBeginDocument{%
  \providecommand\BibTeX{{%
    \normalfont B\kern-0.5em{\scshape i\kern-0.25em b}\kern-0.8em\TeX}}}

\renewcommand\footnotetextcopyrightpermission[1]{} 

\title{VILT: Video Instructions Linking for Complex Tasks}

\author{Sophie Fischer} 
\affiliation{\institution{University of Glasgow}
\city{Glasgow}
\country{Scotland}}
\email{sophie.fischer@glasgow.ac.uk}

\author{Carlos Gemmell} 
\affiliation{\institution{University of Glasgow}
\city{Glasgow}
\country{Scotland, UK}}
\email{c.gemmell.1@research.gla.ac.uk}

\author{Iain Mackie} 
\affiliation{\institution{University of Glasgow}
\city{Glasgow}
\country{Scotland, UK}}
\email{i.mackie.1@research.gla.ac.uk}

\author{Jeffrey Dalton}
\affiliation{\institution{University of Glasgow}
\city{Glasgow}
\country{Scotland, UK}}
\email{jeff.dalton@glasgow.ac.uk}

\renewcommand{\shortauthors}{Sophie Fischer, Carlos Gemmell, Iain Mackie, \& Jeffrey Dalton}

\begin{abstract}

This work addresses challenges in developing conversational assistants that support rich multimodal video interactions to accomplish real-world tasks interactively. 
We introduce the task of automatically linking instructional videos to task steps as ``Video Instructions Linking for Complex Tasks'' (\texttt{VILT}). 
Specifically, we focus on the domain of cooking and empowering users to cook meals interactively with a video-enabled Alexa skill.  
We create a reusable benchmark with 61 queries from recipe tasks and curate a collection of 2,133 instructional ``How-To'' cooking videos.
Studying VILT with state-of-the-art retrieval methods, we find that dense retrieval with ANCE is the most effective, achieving an NDCG@3 of 0.566 and P@1 of 0.644. 
We also conduct a user study that measures the effect of incorporating videos in a real-world task setting, where 10 participants perform several cooking tasks with varying multimodal experimental conditions using a state-of-the-art Alexa TaskBot system.  
The users interacting with manually linked videos said they learned something new 64\% of the time, which is a 9\% increase compared to the automatically linked videos (55\%), indicating that linked video relevance is important for task learning.

\end{abstract}



\begin{document}

\maketitle

\pagestyle{empty}

\section{Introduction}

The first generation of modern assistant systems, such as Alexa, Siri, and Google Assistant was primarily voice-only (with limited touch). However, for complex tasks, previous work shows that instructions are difficult to follow due to their length and complexity \cite{Desai2021HeyGC, Murad2021FindingAN}. Since then, the devices continue to evolve to include screens, cameras, and other sensors. As a result, there is the opportunity to support rich multimodal conversational experiences.

Cooking is an example domain where complex real-world tasks require detailed instructions and guidance to support most users.
However, conversations with most cooking assistants, such as the Amazon Alexa Cooking Skill remain limited. 
For example, recent work finds that it is difficult to explain technically-challenging skills through detailed language \cite{Frummet2019DetectingDI}.
Improving cooking is one of the main focuses of the ongoing Alexa Prize TaskBot Challenge \cite{Gemmell2022_GRILLBot}, the first Alexa challenge to include multimodal elements. As a result, the ability to instruct users and illustrate complex tasks by incorporating video is becoming increasingly important in multimodal dialogue systems.

\begin{figure}
\centering
\includegraphics[width=0.425\textwidth]{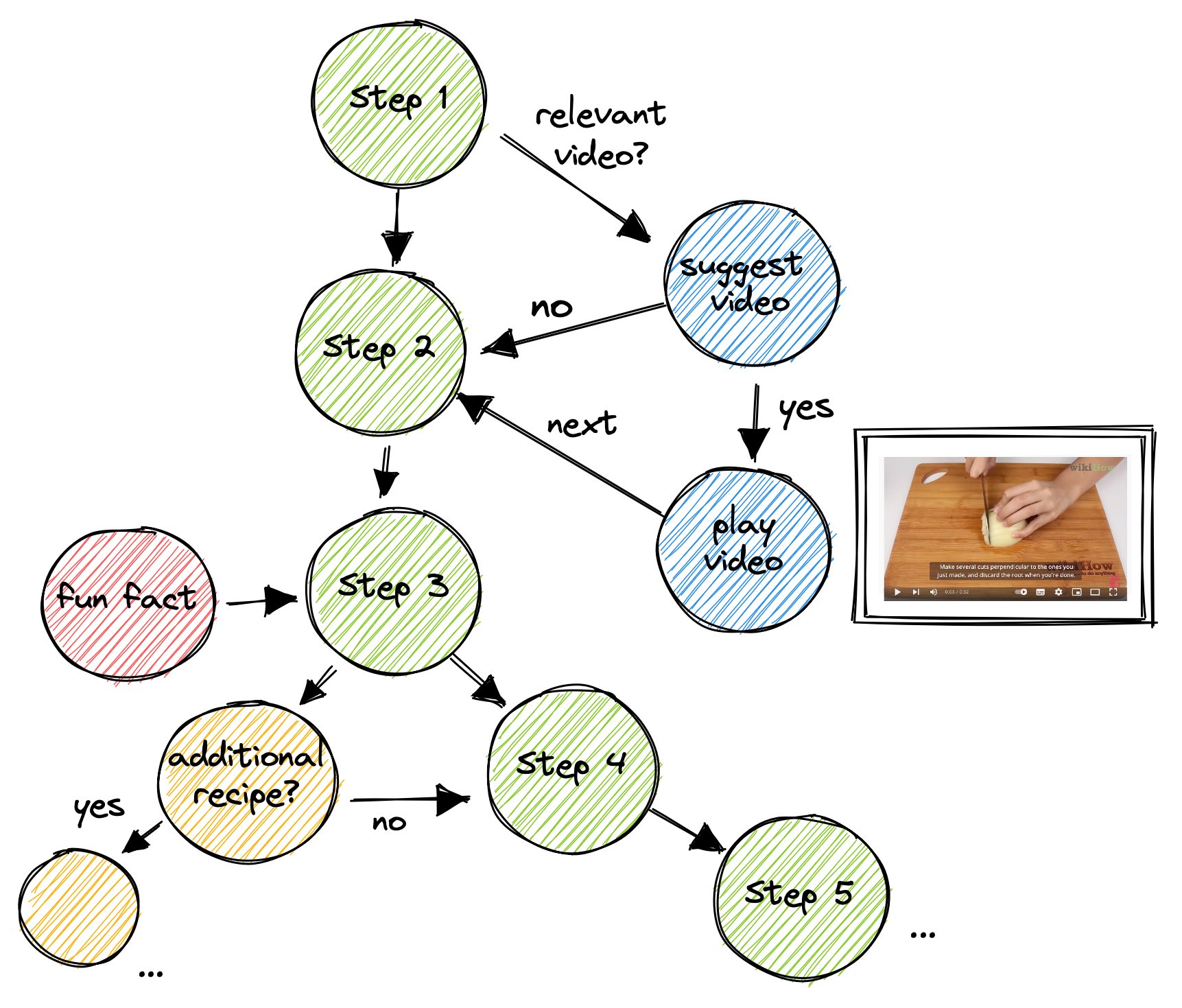}
\caption{
VILT describes linking videos to instructional real-world tasks. For example, a conversational agent can offer a detailed instructional video on ``how to chop an onion'' (blue) to compliment the main cooking steps (green).}
\label{fig:taskgraphs}
\end{figure}

A significant gap in existing task corpora, whether recipes or other tasks, is they are not developed for conversational interfaces and do not contain fine-grained rich multimodal content.
A limited number of tasks have videos for the whole task, and none have fine-grained videos focused on teaching or instructing granular skills (e.g. how to brunoise a carrot). 
As a result, we need new methods for linking video content to task steps appropriate for a rich multimodal conversational experience. 
Figure~\ref{fig:taskgraphs} shows how extra information, in the form of facts, multimodal content or other recipes, can be linked within the standard conversation flow. 

To advance this aim we introduce Video Instructions Linking for Complex Tasks (\texttt{VILT}).
VILT supports introducing instructional video content into a multi-step complex task suitable for a conversational system.
A complex task is defined as a real-world instructional task, and in our case is represented by a recipe that includes more than one cooking method.
An example of a cooking method can be ingredient preparation steps, such as mincing garlic, as well as cooking skills required to complete a recipe step, such as steaming vegetables.
VILT is challenging because the step text is written to inform the user, not to support the linking of general concepts and techniques.
This is demonstrated by the poor BM25 results on VILT based on our evaluation, driven by significant word mismatch that requires more advanced semantic matching.

This work introduces a new benchmark collection of tasks and multimodal video content\footnote{available at https://github.com/grill-lab/VILT}. Furthermore, we create a new task-centric video corpus from Common Crawl and YouTube containing over 2,000 ``How-To'' videos of common cooking methods.
For automatic linking, we frame it as an information retrieval task with queries created from the step instruction descriptions.  
We study the effectiveness of state-of-the-art retrieval methods including sparse retrieval, ANCE \cite{Xiong2021ApproximateNN}, ColBERT \cite{Khattab2020ColBERTEA}, and neural re-ranking  \cite{Nogueira2019PassageRW}.
The results show that ANCE significantly outperforms the other methods and achieves an NDCG@3 of 0.566. The semantic vector similarity overcomes key issues of word mismatch and term weighting when using noisy step text as a query.  
However, the ANCE method only achieves a P@1 score of 0.644, which means approximately a third of the automatically linked videos are irrelevant. 
These results demonstrate that the task is challenging and with significant headroom for improvements.  

Another contribution of this work is that we perform a user study to evaluate the utility of videos during cooking with a conversational agent system in a real task environment.
The experimental setup examines the utility of a screen, testing a voice-only interface and comparing it to an experience with a multimodal experience that incorporates videos.
We also study the differences in experience between automatically linked videos from VILT and those that are manually linked. 
The study includes 10 participants that cook from a pre-selected set of five manually selected tasks. 
It finds that videos are advantageous for helping users accomplish complex cooking tasks.
The test condition with manually linked recipes helps users to feel more empowered and complete complicated tasks successfully.
Similar to the offline VILT experiments, users in the study find that automatically linked videos were less useful than manually selected ones, from 68\% for automatic to 79\% for manual linking.
The most common reason for videos not being useful is that no relevant videos are present in the corpus.

To summarize, this work makes several novel contributions to the study of multimodal conversational agents that incorporate automatically selected instructional content for complex real-world cooking tasks:
\begin{itemize}
    \item We introduce a new benchmark corpus for evaluating the task of automatically linking short instructional videos to fine-grained recipe steps. 
    \item We perform a comparative study of state-of-the-art retrieval methods on VILT, showing that there is significant headroom for improvement.
    \item We conduct a user study that compares automatic and manually selected video content and examines when and how video content is useful in the cooking progress. 
    Specifically, this demonstrates that multimodal video content can result in less confusion, and increases skill learning during task execution.
\end{itemize}

To our knowledge, this is the first work to explore the task of automatically linking video content to ``How-To'' steps for a conversational assistant.
The real-world study shows that videos add significant value to the user experience by decreasing confusion and increasing perceived skill acquisition.
The retrieval experiments find that the task is challenging and that there is significant headroom for future improvement.

\section{Related work}

\citet{Shuster2021MultiModalOD} explore extending conversational agents with images to support multimodal dialogue by leveraging state-of-the-art vision models.
Results suggest that agents that can suggest and converse about images can create a more engaging experience \cite{Shuster2020ImageChatEG}.
Furthermore, many approaches focus on visual question-answering in terms of images \cite{Alayrac2022FlamingoAV, Das2017VisualD} and videos \cite{Le2019MultimodalTN} in goal-oriented conversations, such as selecting and choosing a product.

In the space of video-grounded dialogue, research focuses on understanding video content \cite{Alayrac2016UnsupervisedLF, Malmaud2015WhatsCI}.
However, for conversational experiences, only a few approaches suggest task-specific videos to users for further guidance and engagement.
\citet{Behnke2019AliceID} use manually selected DIY ``How-To'' videos to assist the user learn DIY methods.
In this work, we explore the utility of automatically linked videos with a focus on the cooking domain. 

Previous video corpora include YouCook2, an instructional cooking data corpus containing 2,000 videos \cite{Zhou2018TowardsAL}, and COIN \cite{Tang2021ComprehensiveIV}, consisting of 89 recipes, which are annotated with a series of step descriptions and the corresponding temporal boundaries. Instead of whole recipes, we focus on videos with fundamental techniques that may be relevant to steps in many tasks. 
\citet{Miech2019HowTo100MLA} created a large scale video dataset containing 136 million video clips from ``How-To'' videos. However, Miech at al. remove audio in their videos and compress them, making them unusable for VILT and our user study. 
\citet{Malmaud2015WhatsCI} apply a similar approach to Miech et al., but search the entirety of YouTube for videos tagged with ``cooking'' or ``recipe'', which results in 180,000 videos. However, their video snippets are on average 10 seconds long, which is too short for general techniques.

\section{VILT task}

\label{task definition}

In this section, we define VILT more formally. 
Specifically, \texttt{VILT} is linking instructional videos to steps in a task $T$ with multiple steps $[S_1, ..., S_N]$.
Given a step $S$, we formulate a query $Q$.
For each step query $Q$, we return a relevance-ranked list of video results $[D_1, ..., D_N]$.
We illustrate the task with the example of tabbouleh salad.
This recipe contains the following steps: ``Dice the tomatoes'' ($S_1$), ``Chop parsley and mint leaves'' ($S_2$), ``Cook the bulgur'' ($S_3$) and ``Chop scallions to sprinkle over the salad'' ($S_4$).
For each step $S_1$-$S_4$, there is an underlying cooking technique that the user needs to be able to perform to complete the step successfully.
For each of $S_1$ - $S_4$, we formulate a query $Q$ for which the system needs to retrieve a relevant video $D$.

\subsection{Instructional Video Corpus}
\label{video corpus}

We create a ``How-To'' video corpus using a video processing pipeline to select and download instructional videos from the popular task website, wikiHow.
We retrieve article content from \textit{wikiHow.com} using data from Common Crawl.
We create a wrapper based on BeautifulSoup to extract structured content from the HTML pages.
The wrapper extracts the article title and checks whether the article has a YouTube video on the page.
The result is a candidate set of 18,455 articles with associated instructional videos.

To limit the videos to cooking, we filter the source wikiHow articles to only include those about cooking using a Naive Bayes classifier trained on Reddit threads \cite{Baumgartner2020ThePR} from popular cooking subreddits.
The classifier is applied to the text of the article, resulting in a collection of 2,239 Cooking ``How-To'' articles with their corresponding YouTube videos.
After downloading the final video corpus contains 2,140 videos describing cooking methods.
The representation of the video includes its text, description, and subtitles of the video.
Future work could also consider the automatically extracted objects and actions, and semantic concepts present in the video \cite{Yu2012InformediaE}.
For linking, the video metadata is indexed using Pyserini \cite{Lin2021PyseriniAP}, as described in Section~\ref{indexing}.

\section{VILT annotation and results}

We evaluate how effective current state-of-the-art retrieval methods are at automatically selecting relevant step-specific videos based on step text.
This is challenging because linking must understand the cooking method described in the video and match this to a cooking step, which might also only mention the cooking method implicitly.

\begin{figure}
    \centering
    \includegraphics[width=0.5\textwidth]{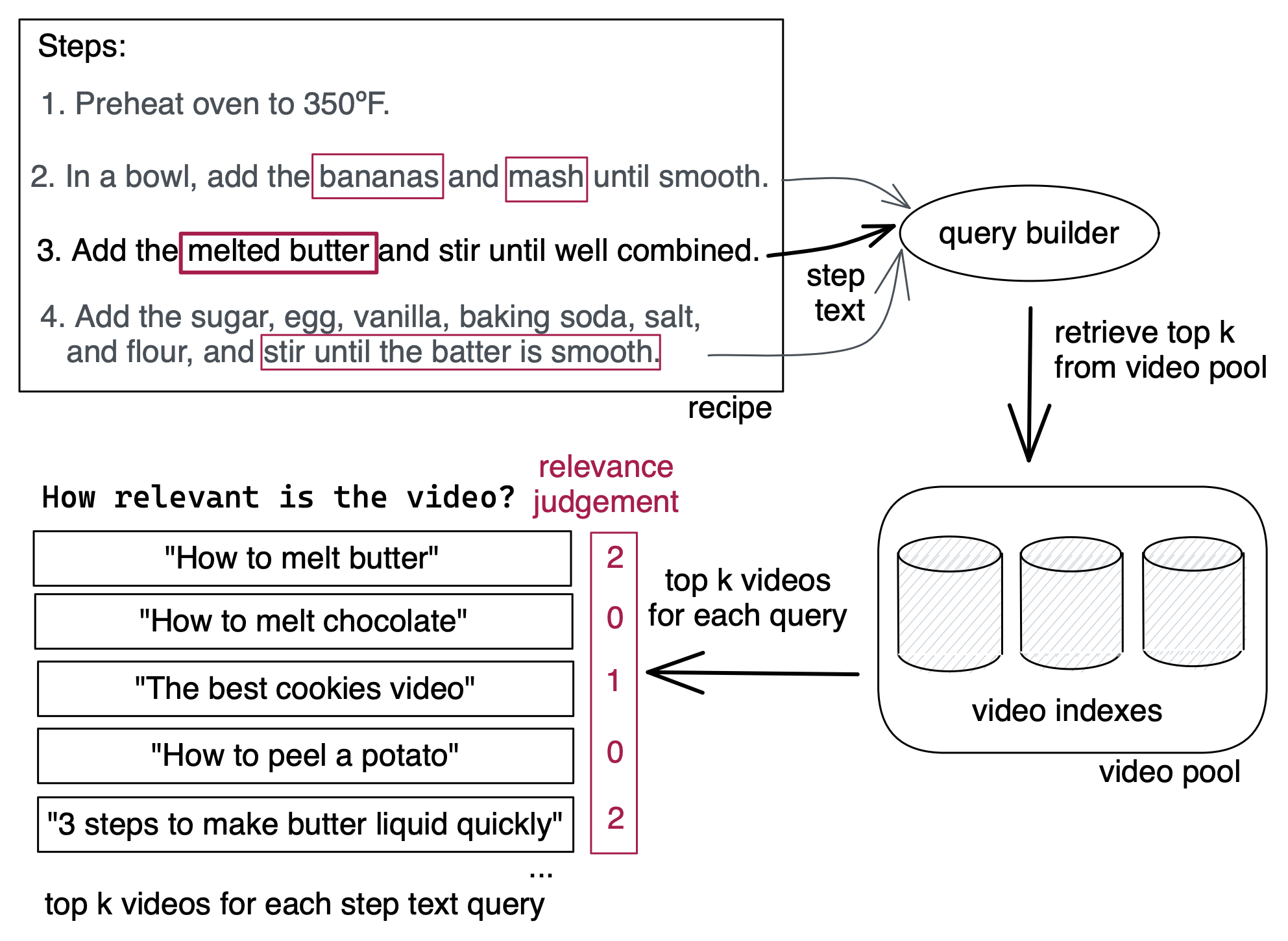}
    \caption{
    Annotation pipeline for relevance judgments. For each recipe component, we retrieve the top \textit{k} videos from the video pool and annotate (\textit{0: not relevant}, \textit{1: somewhat relevant}, \textit{2: very relevant})}
    \label{fig:Annotation pipeline}
\end{figure}

In this section, we detail the creation of the VILT evaluation corpora that consists of tasks and steps (queries), detail the experimental setup, and present the results. 
We note that not all steps have techniques and that some techniques are not covered by the video corpus due to limited scope.

\subsection{Evaluation tasks and process}
\label{taskgraph selection}
\label{annotation guidelines}

To evaluate video linking on cooking tasks, we fetch recipes from \textit{wholefoodsmarket.com} that are present in Common Crawl.
We then convert these recipes into a pre-defined task format with the help of the OAT\footnote{https://github.com/grill-lab/OAT} toolkit.
This results in a collection of 2,275 recipes in the OAT TaskGraph format \cite{Gemmell2022_GRILLBot, Dalton_CISTutorial}.
As part of the annotation pipeline, the annotator can select a task to annotate from a random collection subset in our custom-created annotation tool.
Specifically, the annotation tool splits the task into sections, including requirements text and step text for annotation.

Figure~\ref{fig:Annotation pipeline} shows the annotation pipeline to annotate video content links for a target task.
For a task, the annotator previews the entire list of steps.
For each step, individual queries are built by filtering out stop words and tokenizing the query text.
The annotation tool creates a query and retrieves the top 10 documents from the video pool.
The video pool includes results from different retrieval methods.
Following a similar approach to \citet{mackie2022codec}, documents to annotate are prioritized by (A) taking the top \textit{N} documents from each system run, (B) the presence of the same document across multiple system runs, and (C) randomly sampling the top-ranked documents not annotated.
The annotator is presented the view of the first 10 ranked video documents to assess their relevance on a scale from 0-2 (\textit{0: not relevant}, \textit{1: somewhat relevant}, \textit{2: very relevant}).
A very relevant video (2) describes the cooking method used in the recipe step perfectly, whereas a somewhat relevant video (1) contains information that is useful to the recipe step.
However, this video contains background information about another recipe or talks about different ingredients. 
A video judged as 0 does not contain any relevant information on the cooking method.

\begin{table}
    \centering
    \begin{tabular}{l|l}
    \toprule
    \textbf{Judgement} & \textbf{Video Document Ranking} \\
    \midrule
    {0}     & {580} \\
    {1}    &  {191} \\
    {2} & {60} \\
    {\textbf{TOTAL}} & {\textbf{831}} \\
    \bottomrule
    \end{tabular}
    \caption{Judgment distribution}
    \label{tab:judgements}
\end{table}

The video linking collection includes annotations from 10 tasks, which the annotators chose from a random subset of the collection of 2,275 Wholefoods recipes.
There are linking annotations for 61 query steps across these tasks which contain cooking techniques, chosen from the 189 total recipe steps.
As each method results in approximately 10 videos to annotate, the collection consists of 831 linking judgments.
Table~\ref{tab:judgements} shows the judgment distribution across video document relevance.
It shows that on average there is approximately one very relevant video and approximately three partially relevant videos per evaluation step.

\subsection{Linking retrieval methods}
\label{indexing}

We evaluate our set of test queries against different retrieval methods, such as sparse retrieval (BM25 \cite{Robertson1994SomeSE}), dense retrieval (ANCE \cite{Xiong2021ApproximateNN}, ColBERT \cite{Khattab2020ColBERTEA}) and neural re-ranking (T5 \cite{Nogueira2019PassageRW}).
Each retrieval method retrieves video documents based on the YouTube title and the extracted metadata, including the video description and subtitles.

To compare linking effectiveness, we use standard retrieval measures Mean Average Precision (MAP), Mean Reciprocal Rank (MRR), Precision @ 1, and Normalized Discounted Cumulative Gain at a cutoff of three (NDCG@3).
We focus on precision in the top ranks because typically only one linked video is shown to a user.
We compute all measure using the \textit{ir\_measures} package \cite{MacAvaney2022StreamliningEW}. 

As a baseline in our experiments, we use BM25, tuning parameters $b$ and $k1$ using four folds of cross-validation. 
BM25 is run with the Pyserini framework \cite{Lin2021PyseriniAP}, ANCE and ColBERT use Pyserini's dense retrieval wrappers, and T5 uses PyGaggle.
The T5 neural model is pre-trained using MS Marco \cite{Campos2016MSMA} as part of a two-stage retrieval pipeline with BM25 as the initial retrieval method (BM25+T5).
The ColBERT and ANCE models have been pre-trained on MS MARCO V2.
We evaluate linking by retrieving the top $k=100$ video results for the target task steps.
We use the relevance judgments for the videos (qrels) to calculate retrieval measures on system runs for each query.
To measure significance we use a 5\% paired t-test against the baseline BM25.

\subsection{Results}
\label{ir eval results}

\begin{table}
\centering
\small
\begin{tabular}{lllll}
\toprule
\textbf{} & \textbf{MRR} &  \textbf{NDCG@3} &    \textbf{MAP} & \textbf{P@1}\\
\midrule
\textbf{BM25} & 0.376\phantom{*} &   0.215\phantom{*} & 0.211\phantom{*} & 0.237\phantom{*}\\
\textbf{BM25+T5} & 0.572* &   0.349* & 0.304*  &  0.508* \\
\textbf{ColBERT} & 0.507* &   0.285\phantom{*} & 0.274\phantom{*} & 0.373\phantom{*}\\
\textbf{ANCE} & \textbf{0.735}* &   0.529* & 0.542* & \textbf{0.644*} \\
\bottomrule
\end{tabular}
\caption{Video ranking effectiveness for the annotates test queries. \textbf{Bold} best model and (*) indicates 5\% significance versus BM25.}
\label{tab:IR_eval_methods}
\end{table}

Table~\ref{tab:IR_eval_methods} shows the evaluation of our linking queries. 
We observe that all retrieval methods significantly improve over baseline system BM25 for MRR.
For NDCG@3 and MAP, ANCE and T5 re-ranking perform significantly better than the BM25 baseline.
Based on NDCG@3, ANCE is the best performing method. 

\begin{figure}
    \centering
    \includegraphics[width=0.45\textwidth]{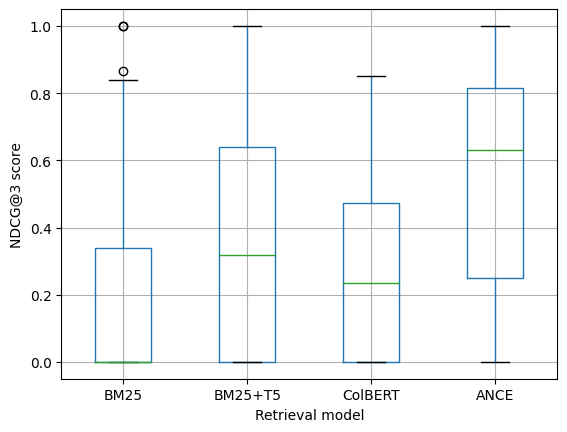}
    \caption{Video ranking effectiveness for Whole Foods tasks with graded judgments.}
    \label{fig:ndcg}
\end{figure}

Figure~\ref{fig:ndcg} shows the effectiveness distribution with graded relevance.
The standard monoT5 model re-ranking on top of BM25 does lead to improvement but still has high variance, which we attribute to noisy step text that is different from typical web query text.
ANCE performs well with a median above 0.6. We attribute this to the dense vector representation being more robust to noise and the sparse lexical word matching. 

We discuss the errors and failures of the approaches in the following section.
For the step ``Grill fillets, skin side down and without flipping, until just cooked through and crisp around the edges, 8 to 10 minutes.'', the relevant instructional video is ``how to grill a fillet of fish''.
However, BM25 retrieves the video ``How To Make A Perfect Burger’’.
This reflects that BM25 matches videos with high lexical overlap to the steps query but does not necessarily match the core concepts effectively.
We attribute this to known issues with long steps being verbose queries \cite{Gupta2015InformationRW}.
In contrast, a dense vector approach from ANCE links the video ``How to Cook whole fish on the BBQ''. 

An example of the ANCE approach failure is  in the ``Greek Lamb Burgers'' task with the step ``Form into 6 patties of even thickness.''
This step is very short, and ambiguous and would require expansion from the task to be effective.
BM25 retrieves a relevant first result with the video ``How To Make A Perfect Burger''. 
In contrast, ANCE retrieves the video ``How to Make Takoyaki'', in which a chef creates makes patties from diced octopus wrapped in dough. This highlights that additional task context would benefit video linking and is a direction for future research.

To focus on precision at the early ranks, ANCE is also the best performing with a P@1 of 0.644, returning at least a partially relevant linking result in approximately two thirds of the task steps.
It also shows that there remains significant headroom for improving the effectiveness. 

\begin{figure}
    \centering
    \includegraphics[width=0.45\textwidth]{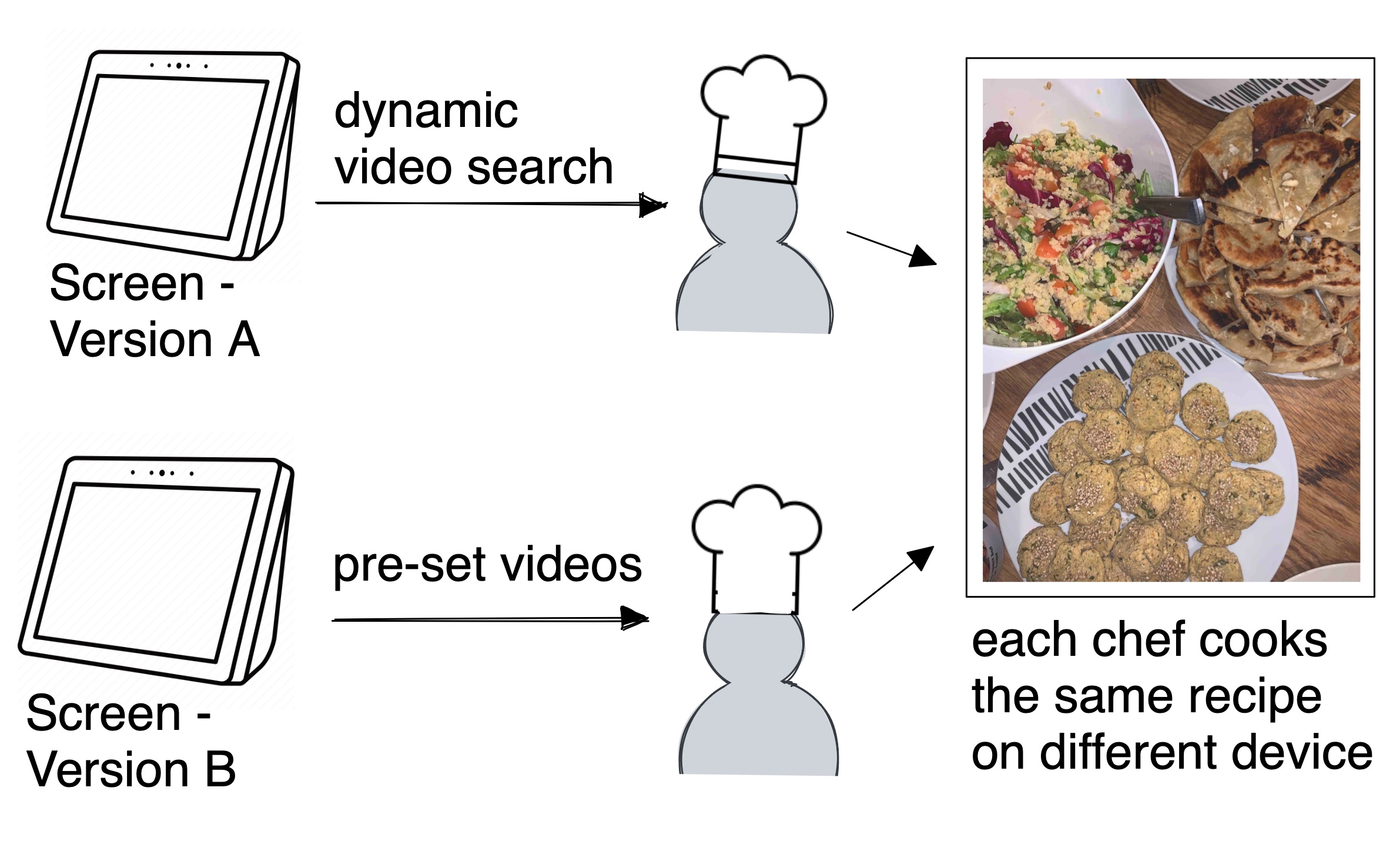}
    \caption{Each recipe is tested with three different setups [screen (dynamic videos), screen (pre-set videos)] in three different experiments.}
    \label{fig:different devices}
\end{figure}

\section{User study}
\label{system eval}
To evaluate the effectiveness of video retrieval in a real-world setting, we perform a user study to show the effects of videos on user satisfaction and engagement.
We test two different versions of the cooking assistant with a screen with linked videos:

\begin{enumerate}
    \item \textbf{Automatic Linking}: We automatically select videos using dynamic retrieval from the corpus with tuned BM25 retrieval.
    \item \textbf{Manual Linking}: We pre-define which video should be shown at what step by manually linking videos.
\end{enumerate}

The setup allows us to study the impact of automatic linking of videos versus a human video selection. 
We predict users will perceive videos linked manually to be more relevant and, as a result, prefer it over the automatic linking method. 

\subsection{Test setup}
\label{system initiative ideas}

To test the video linking, we integrate them into a state-of-the-art Alexa Prize TaskBot system \cite{Gemmell2022_GRILLBot} that is deployed publicly for users in the US.
In previous sections, we glossed over the issue of when and how to show the video content for a task step.
For example, it may be automatically played by the system, it may prompt the user to watch the video, or the user may have to request it.
Based on preliminary experiments, we suggest the video where it is available and prompt the user if they would like to watch it.

To determine when a video should be shown, we use a hand-curated lexicon of common cooking methods, similar to prior work \cite{Malmaud2015WhatsCI}. 
If one is detected in the step, then an appropriate video is selected.
As a form of system initiative, we show videos at the start of the task step performed. 
This decision process and user flow are shown in our original overview Figure~\ref{fig:taskgraphs} where the user is being offered a video matching the recipe step. 

\subsection{User study setup}

We manually select five representative recipes for the user study.
They were selected to form a cohesive meal with multiple dishes that would appeal to users. 
These recipes include a sufficient variety of cooking techniques to have videos with appropriate depth to evaluate the utility of videos for learning a mix of familiar and new skills.

We test the following variables.
Before users start cooking, we ask for their familiarity with the cooking methods used in the recipe.
During the conversation with the agent, we test for the videos watched (for screen systems), the system intents triggered and the conversation length.
After the evaluation, we ask users for their perceived usefulness of the videos and what aspects they learned something about.

The user study consists of two full meals, each of which will be cooked twice with each system version in random order.
Figure~\ref{fig:different devices} shows how each recipe is cooked in different settings.
Meal A consists of tarte flambée and kale salad, and Meal B consists of tabbouleh salad, homemade falafel, and homemade pita bread.

All of our participants are university students between the ages of 20 - 25.
In total, we perform six cooking evenings, with two, three, or four participants in attendance who cook either Meal A or B.
Each cooking evening is conducted in the same kitchen, for a simplified setup.
In total, we conduct 10 separate user conversations, which gives us an insight into how the cooking assistant interacts in a real cooking environment outwith a research setting.

The experiment is structured into three components:
First, users are asked to answer a pre-evaluation survey to gain insights into the user's cooking expertise and the perceived difficulty of the task and cooking methods ahead.
Second, the users attempt the cooking tasks.
Third, they are asked to complete a post-evaluation questionnaire about their perception of the completed cooking task, the usability of the agent, and the usefulness of the videos during the conversation.
In addition to the user survey, we also transcribe the user conversation with the agent for further insights.

\section{Analysis}
\label{user study results}

The user study provides insights into different objective and subjective measures to estimate the effect of videos on user satisfaction, engagement, and their ability to conclude the recipe.
We analyze whether the videos shown helped to increase user satisfaction.
In addition, we evaluate how well automatic linking of videos works compared to manually linked video components.
For all of the videos shown, we estimate how useful and relevant they are to the cooking task.

\subsection{Pre-evaluation survey}

Approximately 70\% of our users had not attempted the recipe before the experiment (68\% ``No'', 28\% ``Yes'', 3\% ``Maybe'').
Furthermore, before starting to cook, we ask the user how comfortable they are with the cooking methods used in the experiment, so that we know if they need to learn the cooking method from the agent.
We observe that across all cooking methods, about 80\% of users feel somewhat or very comfortable about performing this method without additional help.

\begin{figure}[h]
    \centering
    \includegraphics[width=0.45\textwidth]{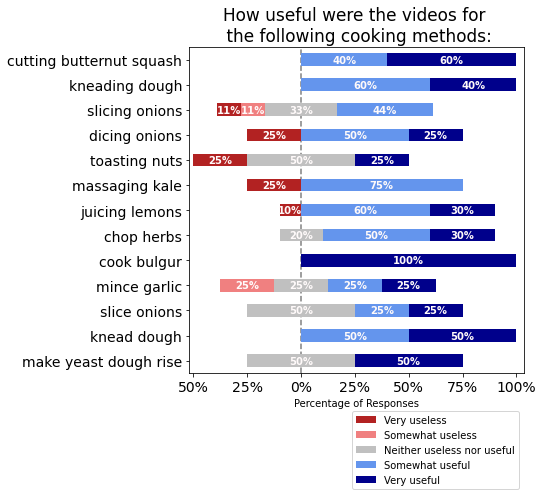}
    \caption{Perceived video utility for each video technique shown across screen devices. Most users agreed that the videos were useful when completing a recipe.}
    \label{fig:likert_scale}
\end{figure}

\subsection{User interactions with videos}

For objective measures, we transcribe conversations held with the agent.
In total, we suggest 35 videos to users across the conversations.
In this set, 14 of the videos were from the manually linked videos.
The users select to watch 7 of the videos (50\%).
For the 21 automatically videos, the user agreed to play 10 of them. 
Although the sample size is small, there was no significant difference in engagement between automatic and manually linked videos.

We ask users how useful the videos were for completing the cooking methods in the post-survey questionnaire.
In Figure~\ref{fig:likert_scale}, we list all 13 cooking methods for which videos were shown by the system.
On average, 68\% of videos were judged as useful when using the system that links videos automatically. 
For the system with the manually pre-selected videos, 79\% of videos were judged as useful.
Furthermore, we ask the user to rate on a scale of 0 - 10 (0 = ``not at all'' to 10 = ``big impact'') how big of a role videos played in their completion of the task.
For automatic, users rated the impact as 4.1 on average, whereas for manual, users rated the impact as 4.8, which also aligns with our findings for perceived video relevance.

\begin{figure}[h]
    \centering
    \includegraphics[width=0.4\textwidth]{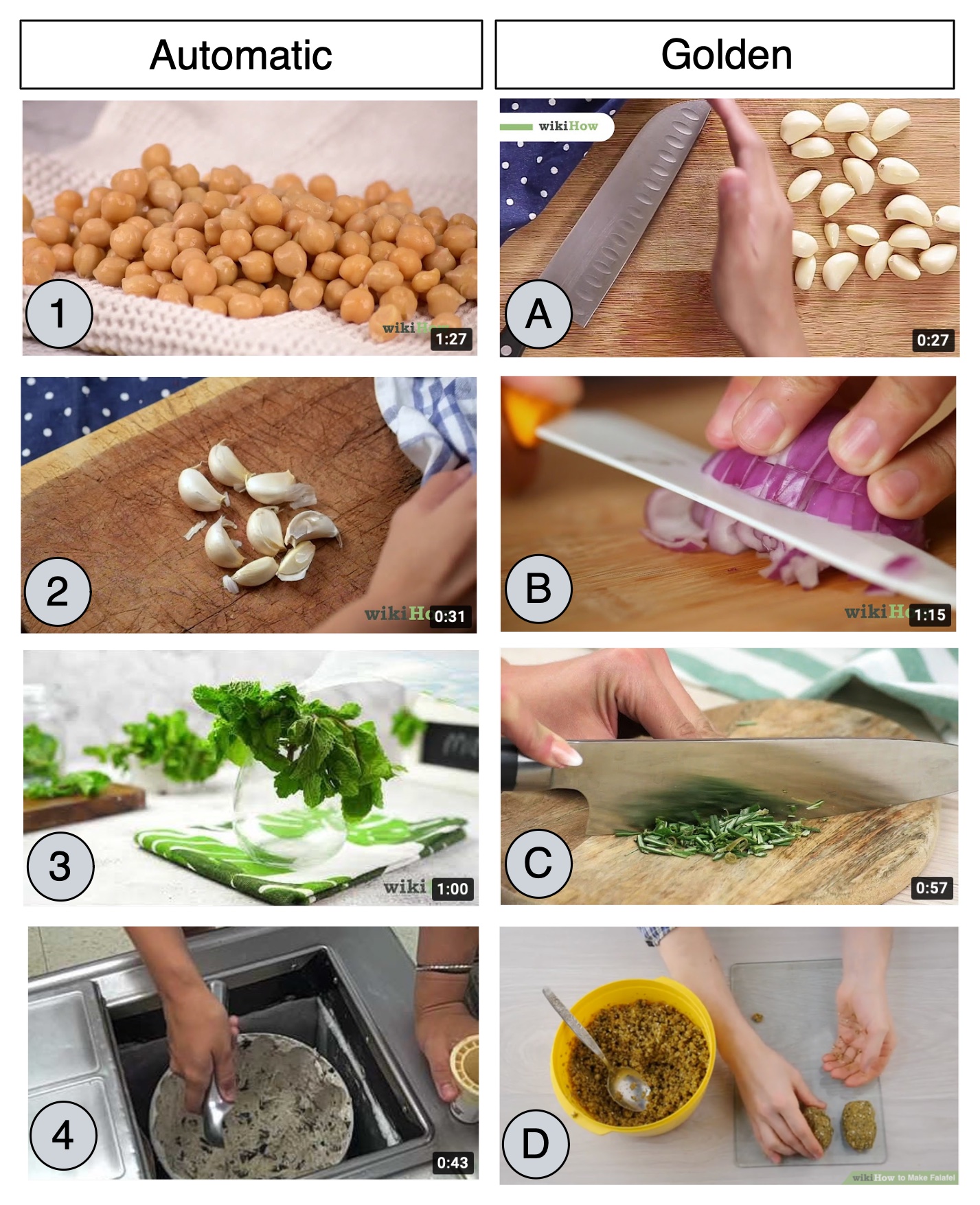}
    \caption{Videos shown in the user study for falafel. Automatic: (1) How to cook falafel, (2) How to peel garlic cloves, (3) How to keep mint leaves fresh, and (4) How to scoop ice cream. Manual: (A) How to mince Garlic, (B) How to dice onions, (C) How to chop herbs, and (D) How to form falafel. Source: WikiHow YouTube.}
    \label{fig:videos_shown}
\end{figure}

Figure~\ref{fig:videos_shown} shows an extract of shown videos for our hand-curated falafel recipe for automatic video retrieval and pre-selected ``manual'' system, which show-cases how the automatic retrieval finds videos that are semantically related, but not always relevant.
For example, the automatic linking finds the video ``How to scoop ice cream'' for the recipe step of ``Using a tablespoon measure, scoop out heaping spoonfuls of the mixture into your hand''.
This recipe step refers to the falafel mixture made in the previous step, which the retrieval system is not able to capture.
This motivates continuing to improve VILT to improve the relevance of videos shown.

Most users state having a high cooking experience for the methods attempted in the user study.
Despite this, for both screen versions, users state they learned something new while interacting with the agent.
For the system with manually linking, users said they learned something new 64\% of the time, which is a 9\% increase compared to the automatic system (55\%). 
When asked what they learned, answers included learning how to cook a recipe in general, but also improvements to the cooking methods that the participants assumed they knew how to do previously.
These results show the potential of VILT to help users engage in tasks and learn new skills.

\section{Conclusions}

\texttt{VILT} introduces the task of linking multimodal video content to steps in complex tasks to support conversational task completion.
We developed a new instructional video corpus including over 2,000 cooking ``How-To'' videos from YouTube.
We evaluate linking with state-of-the-art retrieval methods, including sparse and dense retrieval and neural re-rankers. 
The results show that the dense retrieval ANCE method outperforms other linking approaches. 
However, the Precision@1 score of 0.644 demonstrates significant headroom, with approximately a third of steps not producing a relevant linked video.  

We perform a real-world user study of those cooking to measure the utility of linked videos. Users perceive 68\% of videos shown with the automatic retrieval as somewhat or very useful compared to 79\% with the manually linked content.
It also finds that users report up to a 9\% gain in learning when using a system with manually pre-selected videos compared to a system with automatically linked videos (manual: 64\%, automatic: 55\%).  

Both findings motivate future work on developing new VILT methods and approaches for incorporating videos into conversational agents. 
Specifically, incorporating useful videos at the correct time to empower non-experts to learn new skills and successfully complete complex tasks. 
There are many areas for future work, including new linking approaches, new methods to identify and classify task steps automatically, and the most effective way to incorporate videos to maximize task success rates.

\vskip8pt \noindent
{\bf Acknowledgments.}
The authors would like to thank the rest of the GRILL lab for their on-going support, specifically Federico Rossetto and Paul Owoicho.
This work is supported by the Engineering and Physical Sciences Research Council grant EP/V025708/1. 
It was also supported by the Amazon Alexa Prize TaskBot challenge.

\bibliographystyle{ACM-Reference-Format}
\balance
\bibliography{custom}

\end{document}